\documentclass[pra,onecolumn,showpacs,showkeys,amsmath,nobalancelastpage,floatfix,preprintnumbers,amssymb]{revtex4}

\usepackage[T1]{fontenc}
\usepackage[latin1]{inputenc}
\usepackage[english]{babel}

\usepackage{graphicx}
\usepackage{bm}

\DeclareMathOperator{\trace}{Tr}
\DeclareMathOperator{\Ad}{Ad}
\newcommand{\ket}[1]{\ensuremath{\left| #1 \right \rangle}}
\newcommand{\vect}[1]{\ensuremath{\mathbf{#1}}}
\newcommand{\bind}{bind}
\newcommand{\invar}{\ensuremath{\eta} }

\begin{document}

\title{A new local invariant for quantum gates}
\author{Laura Koponen}
\email{lkoponen@cc.hut.fi}
\author{Ville Bergholm}
\email{vberghol@focus.hut.fi}
\author{Martti\ M.\ Salomaa}
\affiliation{Materials Physics Laboratory, POB 2200 (Technical Physics)\\
FIN-02015 HUT, Helsinki University of Technology, Finland}
\date{\today}

\begin{abstract}
In this paper we study the properties of two-qubit gates.
We review the most common parameterizations for the local equivalence
classes of two-qubit gates and the connections between them.
We then introduce a new discrete local invariant, namely the number of
local degrees of freedom that a gate can \bind{}. The value of this
invariant is calculated analytically for all the local
equivalence classes of two-qubit gates. We find that almost
all two-qubit gates can \bind{} the full six local degrees of freedom
and are in this sense more effective than the controlled-NOT gate
which only can \bind{} four local degrees of freedom.
\end{abstract}


\maketitle

\parindent 0mm
\parskip 5mm



\section{Introduction}

Quantum computation is a novel information processing method in
which classical information is encoded into a quantum-mechanical
system~\cite{nielsen}, called the quantum register.
In most quantum computers the quantum register is a collection of
two-level systems, termed qubits.
The computation is performed by the unitary temporal evolution of the
register, followed by a measurement.
In order to execute a quantum algorithm, one has to be able
to generate the required unitary propagators that are usually referred
to as quantum gates.

It has been shown that almost any fixed two-qubit gate together with
arbitrary single-qubit gates is universal~\cite{dbe,lloyd},
i.e., any $n$-qubit gate may be constructed using only a finite number
of these gates. Conventionally, the elementary gate library is chosen to
consist of the single-qubit rotations $R_x,R_y,R_z$ and the
controlled-NOT gate (CNOT). However, in many realizations, the
CNOT is not the natural choice for the entangling two-qubit gate.
Recently, an optimal construction of an arbitrary two-qubit gate
using three CNOTs and 15 single-qubit rotations has been
introduced~\cite{shende}. In addition, constructions for the
double-CNOT (DCNOT)~\cite{whaley_controlledU}, the controlled-unitary
gates~\cite{whaley_controlledU} and the so-called super controlled
gates~\cite{ye} have been published. A construction using only two
applications of the $B$~gate has been introduced in
Ref.~\cite{whaley_b}, and in Ref.~\cite{yszhang} it is
shown that no other construction using only two applications of a
fixed two-qubit gate exists. Extensions to the $n$-qubit case are
mainly uninvestigated. However, several CNOT-based constructions
with $O(4^n)$ asymptotic behaviour exist, the best of
which~\cite{shende_matrix,bergholm} have CNOT counts of twice the
highest known lower bound~\cite{shende}.

In many of the proposed realizations for quantum computers the
individual qubits are fully controllable, whereas the interqubit
interactions are often fixed.
In addition, single-qubit operations tend to be considerably
faster to implement than multiqubit operations.
This is why it often makes sense to study the \emph{local equivalence
classes} of multiqubit gates instead of the gates themselves.
Two gates are considered equivalent if they can be converted to each
other using only local operations, i.e., tensor products of
single-qubit gates.
The equivalence classes are characterized by local invariants,
which are quantities that are not affected by local operations.

In this paper we briefly review the currently used parameterizations
for the local equivalence classes of two-qubit gates and point out
their equivalence. We then introduce a new discrete local invariant
which describes the number of local degrees of freedom a gate can \emph{\bind{}}.
Finally, we calculate the value of this invariant for all the
local equivalence classes of two-qubit gates.

\section{Local equivalence classes of two-qubit gates}


An $n$-qubit quantum gate $k$ is said to be \emph{local} iff it consists
solely of single-qubit rotations: $k \in SU(2)^{\otimes n} =: \mathbb{L}_n$.
Two $n$-qubit gates $U_1, U_2 \in SU(2^n)$ are said
to be locally equivalent iff $U_2 = k_1 U_1 k_2$, where $k_1, k_2
\in \mathbb{L}_n$. This constitutes an equivalence relation,
which we denote by $U_1 \sim U_2$.

\begin{figure}[h]
\begin{center}
\includegraphics[width=0.55\textwidth]{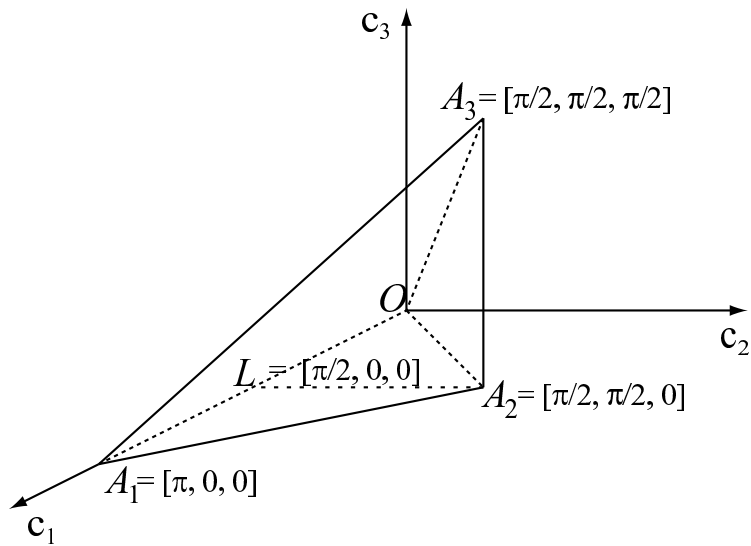}
\caption{\label{fig:weyl}Weyl chamber. Points $O$ and $A_1$
correspond to the identity gate $I$, $A_3$ to the SWAP gate, $L$
to the controlled-NOT gate (CNOT) and $A_2$ to the double
controlled-NOT gate (DCNOT)~\cite{whaley_PRA}.}
\end{center}
\end{figure}

Using the theory of Lie groups
it can be shown~\cite{khaneja_to,whaley_PRA} that any
two-qubit gate $U\in SU(4)$ can be decomposed using the Cartan decomposition as
\begin{equation}
\label{eq:ylexp}
U = k_1 A k_2 =
k_1 \exp \left( \frac{i}{2}(c_1 \sigma_x \otimes \sigma_x
+c_2 \sigma_y \otimes \sigma_y + c_3 \sigma_z \otimes \sigma_z) \right) k_2,
\end{equation}
where $\sigma_{i}$ denote the Pauli matrices,
$k_1, k_2 \in \mathbb{L}_2$ and
$c_1, c_2, c_3 \in \mathbb{R}$. The matrix $A$ is a member of the Cartan
subgroup of the decomposition and carries all the nonlocal properties
of the gate~$U$.
Hence the local equivalence classes of two-qubit gates can be
parameterized by the three scalars
$[c_1, c_2, c_3]$, known as canonical parameters.
This is a minimal set of
parameters since the group $SU(4)$ is 15-dimensional and the local
rotations eliminate
$2 \times \text{dim}(SU(2)^{\otimes 2}) = 12$
degrees of freedom thereof. The canonical parameterization
is visualized in Fig.~\ref{fig:weyl}. The tetrahedron $OA_1 A_2
A_3$ in the figure is called a Weyl chamber. It is defined by the
inequalities $ \pi \ge c_1 \ge c_2 \ge c_3 \ge 0, \pi-c_1 \ge c_2$. The Weyl
chamber contains all the local equivalence classes of two-qubit
gates exactly once, excepting the fact that the triangles $LA_1
A_2$ and $LOA_2$ are equivalent.

The matrix
\begin{equation}\label{eq:quu}
Q = \frac{1}{\sqrt{2}}
\begin{pmatrix}
1 & 0 & 0 & 1 \\
0 & -i & -i & 0 \\
0 & 1 & -1 & 0 \\
-i & 0 & 0 & i
\end{pmatrix}
\end{equation}
is the transformation from the
standard basis of states
$\{\ket{00},\, \ket{01},\, \ket{10},\, \ket{11}\}$
into the Bell basis, also known as the magic basis~\cite{makhlin}.
We use the lower index $B$ to denote the change of basis: $U_B := QUQ^\dagger$.
The magic basis has the special property that local gates expressed
in it are orthogonal.
In other words, conjugation by $Q$ is a group isomorphism between
$SU(2) \otimes SU(2)$ and $SO(4)$. Furthermore, it renders our chosen
Cartan subgroup (generated by
$\sigma_x \otimes \sigma_x$, $\sigma_y \otimes \sigma_y$ and
$\sigma_z \otimes \sigma_z$) diagonal.
These two properties enable us to calculate the canonical parameters
of any given $SU(4)$ gate~$U = k_1 A k_2$. The parameters are obtained from the
spectrum of the matrix $M(U) := U_B^T U_B$ which is given by
\begin{equation} \label{eq:spektri}
\lambda\left(M(U)\right) = \left\{ e^{i(c_1+c_2-c_3)}, e^{i(c_1-c_2+c_3)},
e^{i(-c_1+c_2+c_3)},e^{-i(c_1+c_2+c_3)} \right\}.
\end{equation}
Ref.~\cite{childs} presents an algorithm for extracting the canonical
parameters $c_i$ from this spectrum in a convenient way although it
uses a slightly different notation.
The equivalence of the methods becomes apparent using the equality
$Q^T Q = -\sigma_y\otimes\sigma_y$, since
\begin{equation}
\lambda\left(M(U)\right) = \lambda\left((QUQ^\dagger)^T QUQ^\dagger\right) =
\lambda\left((\sigma_y\otimes\sigma_y)^\dagger U^T
(\sigma_y\otimes\sigma_y) U \right) =
\lambda\left(U (\sigma_y\otimes\sigma_y) U^T (\sigma_y\otimes\sigma_y) \right) =
\lambda(U\tilde{U}).
\end{equation}
Ref.~\cite{shende} presents another system of invariants, namely
the characteristic polynomials $\chi[\gamma_2(U)]$, where
$\gamma_2(U) = U (\sigma_y\otimes\sigma_y) U^T (\sigma_y\otimes\sigma_y)$.
They are completely equivalent to the canonical parameters
since the characteristic polynomial $\chi[\gamma_2(U)]$
carries exactly the same information as 
$\lambda\left(M(U)\right) = \lambda(\gamma_2(U))$.

\begin{figure}[h]
\begin{center}
\includegraphics[width=0.65\textwidth]{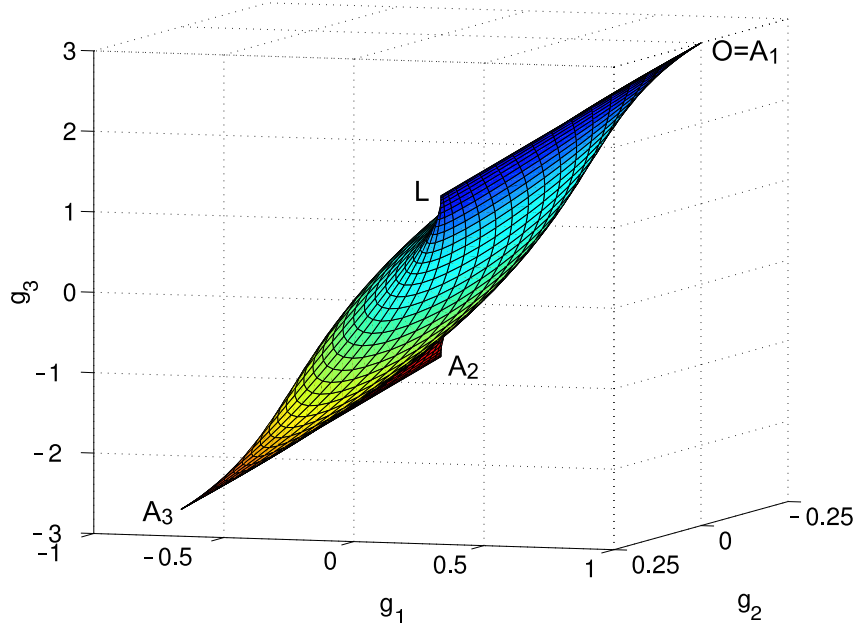}
\caption{\label{fig:rausku}Weyl chamber in the coordinates of the
Makhlin invariants.}
\end{center}
\end{figure}

Another useful parameterization for the two-qubit local equivalence
classes is provided by the Makhlin
invariants $G_1$ and $G_2$~\cite{makhlin}.
For a gate $U\in U(4)$, they are defined as
\begin{equation}\label{eq:makhlinyl}
G_1 = \frac{\trace^2 M(U)}{16 \det U}, \qquad
G_2 = \frac{\trace^2 M(U)-\trace M(U)^2}{4\det U}.
\end{equation}
The Makhlin invariants are by far the easiest ones to calculate.
They, too, provide the same information as the previous invariants
since $\lambda\left(M(U)\right)$ is fully determined by them.
$G_1$~may be complex but $G_2$ is always a real number, which leads to
three real-valued invariants. If $U$ is represented as
in Eq.~(\ref{eq:ylexp}), the Makhlin invariants reduce
to~\cite{whaley_PRA}
\begin{align} \label{eq:makhlin_c}
g_1 &:= \mathrm{Re}\, G_1 = \cos^2 c_1 \cos^2 c_2 \cos^2 c_3-\sin^2
c_1 \sin^2 c_2 \sin^2 c_3, \notag \\
g_2 &:= \mathrm{Im}\, G_1 = \frac{1}{4} \sin 2c_1 \sin 2c_2 \sin 2c_3, \\
g_3 &:= G_2 = 4\cos^2 c_1 \cos^2 c_2 \cos^2 c_3 -4\sin^2 c_1 \sin^2 c_2
\sin^2 c_3 - \cos 2c_1 \cos 2c_2 \cos 2c_3 \notag.
\end{align}
Example values of the invariants of different gates are given in
Table~\ref{table:makhlinit}.
The set of all the two-qubit gate
equivalence classes in the Makhlin parameter space is presented in
Fig.~\ref{fig:rausku}. The surface is given by the equations
\begin{align}
g_1 &= \cos^2 s \cos^4 t - \sin^2 s \sin^4 t \notag \\
g_2 &= \frac{1}{4} \sin(2s) \sin^2(2t) \\
g_3 &= 4 g_1 - \cos(2s) \cos^2(2t) \notag,
\end{align}
where $s \in [0,\pi], \, t \in [0,\pi/2]$. The surface and the
inside of the object correspond to the surface and the inside of
the Weyl chamber, respectively.

\begin{table}
\begin{tabular}{l|ccc|ccc}
\hline
Gate & $c_1$ & $c_2$ & $c_3$ & $g_1$ & $g_2$ & $g_3$ \\
\hline
$I$     & 0 & 0 & 0      & 1  & 0 & 3  \\
SWAP    & $\frac{\pi}{2}$ & $\frac{\pi}{2}$ &  $\frac{\pi}{2}$     & -1 & 0 & -3 \\
CNOT    & $\frac{\pi}{2}$ & 0 & 0      & 0  & 0 & 1  \\
DCNOT   & $\frac{\pi}{2}$ & $\frac{\pi}{2}$ & 0 & 0 & 0   & -1\\
$\sqrt{\text{SWAP}}$ & $\frac{\pi}{4}$ & $\frac{\pi}{4}$ & $\frac{\pi}{4}$   & 0 & $\frac{1}{4}$ & 0 \\
$\sqrt{\text{SWAP}}^{\,-1}$  & $\frac{3\pi}{4}$ & $\frac{\pi}{4}$ & $\frac{\pi}{4}$   & 0 & -$\frac{1}{4}$ & 0 \\
B       & $\frac{\pi}{2}$ & $\frac{\pi}{4}$ & 0    & 0 & 0 & 0 \\
controlled-U & $\alpha$ & 0 & 0 & $\cos^2(\alpha)$ & 0 & $2\cos^2(\alpha)+1$ \\
SPE     &  $\frac{\pi}{2}$ & $\alpha$ & 0 & 0 & 0 & $\cos(2\alpha)$\\
\hline
\end{tabular}
\caption{\label{table:makhlinit} Values of the canonical and Makhlin
invariants for some common gates. SPE denotes a special perfect entangler~\cite{rezakhani,ye}.}
\end{table}

\section{The local invariant \invar}


Let us use
\begin{equation}\label{eq:local_rot_n}
L^k_n(\vect{a}, \vec{\theta}) :=
\exp \left( \sum_{j=1}^{3n} a_j(\vec{\theta}) X_j \right), \qquad
L^k_n(\vect{a}, \vec{\theta}) \in \mathbb{L}_n \quad \forall \vec{\theta} \in \mathbb{R}^k,
\end{equation}
where $j$ runs over the $3n$ local
generators of $SU(2^n)$, to denote a $k$-parameter family of $n$-qubit
local gates. It is defined by the function $\vect{a}: \mathbb{R}^k \to
\mathbb{R}^{3n}$.
The generators $X_j$ are normalized such that they are orthonormal
with respect to the inner product
$\left\langle X, Y \right\rangle := \trace \left( X^\dagger Y \right)$.

A gate $U \in SU(2^n)$ is said to \emph{leak} $k$~local degrees of
freedom iff there exist nondegenerate functions $\vect{a}$ and
$\vect{b}$ such that
\begin{equation} \label{eq:pass}
U L^k_n(\vect{a}, \vec{\theta}) = L^k_n(\vect{b}, \vec{\theta}) U \quad
\forall \vec{\theta} \in \mathbb{R}^k.
\end{equation}
A gate \bind s the
local degrees of freedom that it does not leak. We define a
function \invar: $SU(2^n) \to \mathbb{N}$ to indicate the number of
local degrees of freedom that an $n$-qubit gate $U$ binds. We
always have $\max \invar \le 3n$, i.e., at most three degrees of
freedom for each qubit.

Assume now that the functions $\vect{a}$ and $\vect{b}$
satisfy Eq.~(\ref{eq:pass}) for the gate $U$.
For a gate $V = k_1 U k_2$, where $k_1, k_2 \in \mathbb{L}_n$,
we obtain
\begin{equation}
\label{vjohto}
V \: \left[ k_2^{\dagger} L^k_n(\vect{a}, \vec{\theta}) k_2 \right]
= k_1 U k_2 \: k_2^{\dagger} L^k_n(\vect{a}, \vec{\theta})  k_2
= k_1 L^k_n(\vect{b}, \vec{\theta}) \: U k_2
= \left[ k_1 L^k_n(\vect{b}, \vec{\theta}) k_1^{\dagger} \right] \: V.
\end{equation}
We also have
\begin{align}
k_1 L^k_n(\vect{b}, \vec{\theta}) k_1^{\dagger}
= \exp \left( \sum_{j=1}^{3n} b_j(\vec{\theta}) \Ad(k_1) X_j \right)
= \exp \left( \sum_{j=1}^{3n} \tilde{b}_j(\vec{\theta}) X_j \right)
=  L^k_n(\tilde{\vect{b}}, \vec{\theta}),
\end{align}
since $\Ad(g)$ is a linear bijection and $k_1$ is a local gate.
If $\vect{b}$ is nondegenerate then so is $\tilde{\vect{b}}$.
A similar argument naturally holds for
$ k_2^{\dagger} L^k_n(\vect{a}, \vec{\theta}) k_2$, which yields
$V L^k_n(\tilde{\vect{a}}, \vec{\theta})
L^k_n(\tilde{\vect{b}}, \vec{\theta}) V$
and proves that \invar is indeed a local invariant.

Equation~\eqref{eq:pass} is equivalent to
\begin{align}
U \exp \left( \sum_{j=1}^{3n} a_j(\vec{\theta}) X_j \right) U^{\dagger}
= \exp \left( \sum_{j=1}^{3n} a_j(\vec{\theta}) \Ad(U) X_j \right)
= \exp \left( \sum_{k=1}^{3n} b_k(\vec{\theta}) X_k \right)
\end{align}
This is fulfilled if
\begin{align} \label{eq:der}
\sum_{j=1}^{3n} a_j(\vec{\theta}) \Ad(U) X_j
= \sum_{k=1}^{3n} b_k(\vec{\theta}) X_k.
\end{align}
Now, as we take a sidewise inner product $\langle \cdot, X_i \rangle$
with each of the $4^n-1$ generators of~$SU(2^n)$, we obtain equivalently
\begin{equation}
\sum_{j=1}^{3n} W_{ij} a_j(\vec{\theta})
= \sum_{k=1}^{3n} b_k(\vec{\theta}) \delta_{ki}, \qquad i=1, 2, \ldots, 4^n-1,
\end{equation}
where $W_{ij} = \trace\left( U X_j^{\dagger} U^{\dagger} X_i \right)$.
The generators $X_j$ are antihermitian and $U$ is unitary. This
implies that the elements $W_{ij}$ are real. Written in
matrix form this is
\begin{equation}\label{eq:yhtryhm}
W \vect{a}(\vec{\theta}) =
\begin{pmatrix}
W_L \\ W_N
\end{pmatrix}
\vect{a}(\vec{\theta}) =
\begin{pmatrix}
\vect{b}(\vec{\theta}) \\ \vec{0}
\end{pmatrix} \quad \forall \vec{\theta} \in \mathbb{R}^k,
\end{equation}
where $W_L \in \mathbb{R}^{3n \times 3n}$,
$W_N \in \mathbb{R}^{(4^n - 1 - 3n) \times 3n}$
and the indices $L$ and $N$ stand for local and nonlocal,
respectively. Hence, we must have
$\vect{a}(\vec{\theta}) \in \ker{W_N}$ for all values of
$\vec{\theta}$. Moreover, since $\vect{b}$ must have the same
dimensionality as $\vect{a}$, the component of $\ker W_N$ parallel to
$\ker W_L$ must be disregarded. Using the rank-nullity theorem we
finally obtain
\begin{equation}
\invar(U) = 3n -\dim(\ker W_N) +\dim(\ker W_L  \cap \ker W_N).
\end{equation}

\section{\invar for two-qubit gates}

For the set of two-qubit gates $U \in SU(4)$, $\max \invar \le 6$.
It is obvious that $\invar(\text{I}) = 0$ and $\invar(\text{SWAP}) = 0$
since all local gates and hence all local degrees of freedom may be
commuted through these gates. It is also known that
$\invar(\text{CNOT}) = 4$ and $\invar(\text{DCNOT}) = 4$.
The result for CNOT is obtained by combining the commutation
properties of CNOT with the Euler rotations $R_z$ and $R_x$ and the
fact that an arbitrary two-qubit gate may be implemented using at most
three CNOTs~\cite{shende,bullock23,vidal,vatan2}.
Similar arguments for the DCNOT are presented in
Ref.~\cite{whaley_controlledU}, including the explicit implementation
of an arbitrary two-qubit gate using three DCNOTs.
Also, from the construction of Ref.~\cite{whaley_b}, it is clear that
$\invar(B) \ge 5$.
Apart from such observations, no explicit calculations for \invar have
been presented in the literature so far.

We will now proceed to derive an analytical expression for $\invar$
for an arbitrary two-qubit gate.
Because \invar is a local invariant, it is enough to consider
gates of the type
\begin{equation}
\label{eq:nonlocalgate}
A = \exp \left( \frac{i}{2}(c_1 \sigma_x \otimes \sigma_x
+ c_2 \sigma_y \otimes \sigma_y + c_3 \sigma_z \otimes \sigma_z)
\right) =
\exp \bigg( \frac{i}{2}
\begin{pmatrix}
c_3 & 0 & 0 & c_1-c_2  \\
0 & -c_3 & c_1+c_2 & 0 \\
0 & c_1+c_2 &-c_3 & 0  \\
c_1-c_2 & 0 & 0 & c_3
\end{pmatrix}
\bigg)
\end{equation}
which represent all the nonlocal equivalence classes.
The calculation of the elements of $W_L$ and $W_N$
is straightforward. Calculating the matrix exponential and
simplifying the expression using elementary trigonometric
identities results in
\begin{equation}\label{eq:mnmatr}
W_L = \begin{pmatrix}
l^1_{1,1} & 0 & 0 & l^1_{1,2} & 0 & 0 \\
0 & l^2_{1,1} & 0 & 0 & l^2_{1,2} & 0 \\
0 & 0 & l^3_{1,1} & 0 & 0 & l^3_{1,2} \\
l^1_{2,1} & 0 & 0 & l^1_{2,2} & 0 & 0 \\
0 & l^2_{2,1} & 0 & 0 & l^2_{2,2} & 0 \\
0 & 0 & l^3_{2,1} & 0 & 0 & l^3_{2,2}
\end{pmatrix}, \qquad
W_N = \begin{pmatrix}
0 & 0 & 0 & 0 & 0 & 0 \\
0 & 0 & n^3_{1,1} & 0 & 0 & n^3_{1,2} \\
0 & n^2_{1,1} & 0 & 0 & n^2_{1,2} & 0 \\
0 & 0 & n^3_{2,1} & 0 & 0 & n^3_{2,2} \\
0 & 0 & 0 & 0 & 0 & 0 \\
n^1_{1,1} & 0 & 0 & n^1_{1,2} & 0 & 0 \\
0 & n^2_{2,1} & 0 & 0 & n^2_{2,2} & 0 \\
n^1_{2,1} & 0 & 0 & n^1_{2,2} & 0 & 0 \\
0 & 0 & 0 & 0 & 0 & 0 \\
\end{pmatrix},
\end{equation}
where the non-zero elements are
\begin{align}\label{eq:mnsel}
L^1 &:=
\begin{pmatrix}
l^1_{1,1} & l^1_{1,2} \\
l^1_{2,1} & l^1_{2,2}
\end{pmatrix} =
\begin{pmatrix}
\cos c_2 \cos c_3 & \sin c_2 \sin c_3 \\
\sin c_2 \sin c_3 & \cos c_2 \cos c_3
\end{pmatrix}, \qquad
N^1 &:=
\begin{pmatrix}
n^1_{1,1} & n^1_{1,2} \\
n^1_{2,1} & n^1_{2,2}
\end{pmatrix} =
\begin{pmatrix}
\sin c_2 \cos c_3 & -\cos c_2 \sin c_3 \\
-\cos c_2 \sin c_3 & \sin c_2 \cos c_3
\end{pmatrix}, \notag \\
L^2 &:=
\begin{pmatrix}
l^2_{1,1} & l^2_{1,2} \\
l^2_{2,1} & l^2_{2,2}
\end{pmatrix} =
\begin{pmatrix}
\cos c_1 \cos c_3 & \sin c_1 \sin c_3 \\
\sin c_1 \sin c_3 & \cos c_1 \cos c_3
\end{pmatrix}, \qquad
N^2 &:=
\begin{pmatrix}
n^2_{1,1} & n^2_{1,2} \\
n^2_{2,1} & n^2_{2,2}
\end{pmatrix} =
\begin{pmatrix}
-\sin c_1 \cos c_3 & \cos c_1 \sin c_3 \\
\cos c_1 \sin c_3 & -\sin c_1 \cos c_3
\end{pmatrix}, \notag\\
L^3 &:=
\begin{pmatrix}
l^3_{1,1} & l^3_{1,2} \\
l^3_{2,1} & l^3_{2,2}
\end{pmatrix} =
\begin{pmatrix}
\cos c_1 \cos c_2 & \sin c_1 \sin c_2 \\
\sin c_1 \sin c_2 & \cos c_1 \cos c_2
\end{pmatrix}, \qquad
N^3 &:=
\begin{pmatrix}
n^3_{1,1} & n^3_{1,2} \\
n^3_{2,1} & n^3_{2,2}
\end{pmatrix} =
\begin{pmatrix}
\sin c_1 \cos c_2 & -\cos c_1 \sin c_2 \\
-\cos c_1 \sin c_2 & \sin c_1 \cos c_2
\end{pmatrix}.
\end{align}

From Eqs.~\eqref{eq:mnmatr}--\eqref{eq:mnsel} it is seen that
Eq.~\eqref{eq:yhtryhm} decomposes into six
separate equations:
\begin{equation}
L^i \begin{pmatrix} a_i \\ a_{i+3} \end{pmatrix}
= \begin{pmatrix} b_i \\ b_{i+3} \end{pmatrix}, \qquad
N^i \begin{pmatrix} a_i \\ a_{i+3} \end{pmatrix}
= \begin{pmatrix} 0 \\ 0 \end{pmatrix}, \qquad
i=1,2,3.
\end{equation}
Each block $L^i$ produces a two-dimensional null space iff all the
elements of $L^i$ equal zero. A one-dimensional null space is
formed iff $\det L^i = \cos(c_j+c_k) \cos(c_j-c_k) = 0$, where
$\epsilon_{ijk} = 1$, but $N^i \ne 0$.
Similarly,
each block $N^i$ produces a two-dimensional null space iff all the
elements of $N^i$ equal zero, and a one-dimensional null space iff
$\det N^i = (-1)^{i+1} \sin (c_j+c_k)\sin (c_j-c_k) = 0$, where
$\epsilon_{ijk} = 1$, but $N^i \ne 0$.

Taking into account the correlations among the elements of the
matrices $L^i$ and $N^i$ we find that in the two-qubit case
$\ker W_L \cap \ker W_N = \{ \vec{0} \}$ always.
Thus we have $\invar = 6 -\dim(\ker W_N)$ and the number of local
degrees of freedom leaked is given by the nullity of $W_N$.
The results for all the possible values
of $[c_1, c_2, c_3]$ are collected in Table~\ref{table:tulos}.
One notices that everywhere inside the Weyl chamber \invar reaches
its maximum value of 6. At the vertices $O=A_1$ and $A_3$
$\invar = 0$, on the edges between them $\invar = 3$, on the edges
$OA_1, A_2A_3$ $\invar = 4$ and on the faces $OA_1A_3, OA_2A_3, A_1A_2A_3$
$\invar = 5$.

\begin{table}
\begin{tabular}{llc}
\\
\hline
$[c_1, c_2, c_3]$ & Set in the Weyl chamber & \invar\\
\hline
$[0,0,0]\stackrel{\wedge}{=}[\pi,0,0]$  & $O,A_1$                                & 0\\
$[\pi/2,\pi/2,\pi/2]$                    & $A_3$                                 & 0\\
$[x,x,x], x\ne 0, x\ne \pi/2$           & $OA_3\setminus \{O,A_3\}$              & 3\\
$[\pi -x,x,x], x\ne 0, x\ne\pi/2$       & $A_1A_3\setminus \{A_1,A_3\}$          & 3\\
$[x,0,0]\stackrel{\wedge}{=}[\pi -x,0,0], x\ne 0$ & $OA_1\setminus \{O,A_1\}$    & 4\\
$[\pi/2,\pi/2,x],x \ne \pi/2$           & $A_2A_3\setminus \{A_3\}$              & 4\\
$[x,x,y],x\ne y,x \ne \pi/2$            & $OA_1A_3\setminus \{OA_3,A_1A_3\}$     & 5\\
$[x,y,y],x\ne y, x+y \ne \pi, y \ne 0$  & $OA_2A_3\setminus \{OA_3,A_2A_3\}$     & 5\\
$[\pi -x,x,y],x\ne y,x \ne \pi/2$       & $A_1A_2A_3\setminus \{A_1A_3,A_2A_3\}$ & 5\\
$\{$All other points$\}$
& $\{$All other points$\}$               & 6\\
\hline
\end{tabular}
\caption{\label{table:tulos} $\invar$, or the number of local degrees of
freedom bound, for the local equivalence
classes of two-qubit gates.}
\end{table}

The number of local degrees of freedom that the gate $U$ leaks is
obtained as the number of pairs of equal eigenvalues $\lambda_i$
in the spectrum of the matrix $M(U)$, presented in Eq.~\eqref{eq:spektri}.
In other words, any $n$-fold eigenvalue of
$M(U)$ indicates $n(n-1)/2$ local degrees of freedom that pass through
the gate $U$. Translated to the language of the Weyl chamber, each
Weyl symmetry plane the point $[c_1, c_2, c_3]$ touches causes the
gate to leak one local degree of freedom.



\section{Conclusion}

In this paper we have introduced a new local invariant \invar{}
for quantum gates, indicating the number of local degrees of freedom
a gate can bind.
Furthermore, we have analytically calculated the value of this
invariant for all two-qubit gates. We have found that almost all
two-qubit gates can bind the full six local degrees of freedom.
However, most of the commonly occurring gates such as CNOT or
$\sqrt{\text{SWAP}}$ are exceptions to the rule, performing much worse
in this sense. 

The meaning of \invar{} is illustrated by considering the lower bounds on gate counts
for a generic $n$-qubit circuit. Let the gate library consist of
all one-qubit gates and a fixed two-qubit gate $U$.
Then almost all $n$-qubit gates cannot be
simulated with a circuit consisting of fewer than 
\begin{equation}
N_U = \left\lceil \frac{4^n-3n-1}{\invar(U)} \right\rceil
\end{equation}
applications of the two-qubit gate. This result is a straightforward
generalization of Proposition III.1 in Ref.~\cite{shende}. The gates 
binding the full six degrees of freedom are thus expected to be the
most efficient building blocks for multiqubit gates.

\begin{acknowledgments}
This research is supported by the Academy of Finland (project No.
206457). VB thanks the Finnish Cultural Foundation for financial
support.

In memoriam Prof. Martti M. Salomaa (1949-2004)
\end{acknowledgments}

\appendix

\section{Mathematical prerequisites}


The Lie algebra $\mathfrak{g}$ of a linear Lie group
$G < \text{GL}(n, \mathbb{K})$ is the set
\begin{equation}
\mathfrak{g} := \{X \in \mathbb{K}^{n \times n} |
\exp(t X) \in G \quad \forall t \in \mathbb{R} \}.
\end{equation}
In can be shown that~$\mathfrak{g}$ is a real vector space spanned by
the generators of~$G$. For example, the Lie
algebra $\mathfrak{su}(n)$ of the group $\text{SU}(n)$ consists
of all the $n \times n$ complex antihermitian traceless matrices.

The adjoint representation of a Lie group~$G$, $\Ad: G \to
\text{Aut}(\mathfrak{g})$, is a group homomorphism defined by
\begin{equation}
\Ad(g) X := g X g^{-1} \quad (g \in G, X \in \mathfrak{g}).
\end{equation}
It behaves in a rather simple way in exponentiation:
\begin{equation}
\exp\left( \Ad(g) X \right) = g \exp\left( X \right) g^{-1}
\quad \text{for all } g \in G, \: X \in \mathfrak{g}.
\end{equation}
Also, if we define an inner product
$\left\langle X, Y \right\rangle := \trace \left( X^\dagger Y \right)$
for~$\mathfrak{g}$, we find that it is preserved by the adjoint representation:
\begin{equation}
\left\langle \Ad(g) X, \Ad(g) Y \right\rangle
= \left\langle X, Y \right\rangle \quad \text{for all } g \in G, \: X,Y \in \mathfrak{g}.
\end{equation}
As a concrete example, the adjoint representation keeps orthonormal
bases of $\mathfrak{g}$ orthonormal.
\bibliography{weyl}

\end{document}